\documentclass{iau_FM}
\usepackage{graphicx}
\usepackage{float}
\usepackage{wrapfig}
\usepackage{natbib}
\usepackage{listings}
\usepackage[colorlinks=true, allcolors=blue]{hyperref} 
\usepackage{xcolor}

\definecolor{codegreen}{rgb}{0,0.6,0}
\definecolor{codegray}{rgb}{0.5,0.5,0.5}
\definecolor{codepurple}{rgb}{0.58,0,0.82}
\definecolor{backcolour}{rgb}{0.95,0.95,0.92}

\lstdefinestyle{astrophysix_style}{
    backgroundcolor=\color{backcolour},   
    commentstyle=\color{codegreen},
    keywordstyle=\color{magenta},
    numberstyle=\tiny\color{codegray},
    stringstyle=\color{codepurple},
    basicstyle=\ttfamily\footnotesize,
    breakatwhitespace=false,         
    breaklines=true,                 
    captionpos=b,                    
    keepspaces=true,                 
    numbers=left,                    
    numbersep=5pt,                  
    showspaces=false,                
    showstringspaces=false,
    showtabs=false,                  
    tabsize=2
}

\title[The Galactica open simulation database] 
{The Galactica database: an open, generic and versatile tool for the dissemination of simulation data in astrophysics}

\author[Damien Chapon \& Patrick Hennebelle]   
{Damien Chapon$^1$
 \and Patrick Hennebelle$^2$}
   
\affiliation{$^1$IRFU, CEA, Université Paris-Saclay, F-91191 Gif-sur-Yvette, France\\ email: {\tt damien.chapon@cea.fr} \\[\affilskip]
$^2$Universit\'e Paris Diderot, AIM, Sorbonne Paris Cit\'e, CEA, CNRS, F-91191 Gif-sur-Yvette, France\\email: {\tt patrick.hennebelle@cea.fr}}

\pubyear{2024}
\jname{Astronomy in Focus, Focus Meeting 7} 
\editors{Diana M.~Worrall, ed.}
\begin{document}

\maketitle

\begin{abstract}

The Galactica simulation database is a platform designed to assist computational astrophysicists with their open science approach based on FAIR (Findable, Accessible, Interoperable, Reusable) principles. It offers the means to publish their numerical simulation projects, whatever their field of application or research theme and provides access to reduced datasets and object catalogs online. The application implements the Simulation Datamodel IVOA standard.

To provide the scientific community indirect access to raw simulation data, Galactica can generate, on an "on-demand" basis, custom high-level data products to meet specific user requirements. These data products, accessible through online WebServices, are produced remotely from the raw simulation datasets. To that end, the Galactica central web application communicates with a high-scalability ecosystem of data-processing servers called \textit{Terminus} by means of an industry-proven asynchronous task management system. Each \textit{Terminus} node, hosted in a research institute, a regional or national supercomputing facility, contributes to the ecosystem by providing both the storage and the computational resources required to store the massive simulation datasets and post-process them to create the data products requested on Galactica, hence guaranteeing fine-grained sovereignty over data and resources.

This distributed architecture is very versatile, it can be interfaced with any kind of data-processing software, written in any language, handling raw data produced by every type of simulation code used in the field of computational astrophysics. Its generality and versatility, together with its excellent scalability makes it a powerful tool for the scientific community to disseminate numerical models in astrophysics in the exascale era.

\keywords{database, numerical simulation, open science, computational astrophysics, FAIR principles}


\end{abstract}

\firstsection 
\section{Introduction}\label{sec_intro}

The Amsterdam call for Open Science in 2016 started to promote public access to both the scientific publications and the data obtained with public funds.
This call has been implemented at national level all over the world in recent years in the form of Open Science plans and programs to encourage the effective sharing of publications and research data. The first international framework on open science, the UNESCO Recommendation on Open Science \citep{UNESCO2021}, was adopted by 193 countries attending UNESCO’s General Conference in 2021.

In recent years, data publication and reuse has become a key requirement demanded by both funding agencies and resource (computation or observation time) allocation committees.

In the field of astrophysics, a number of initiatives have been taken to disseminate scientific data, mostly in astronomy and to a lesser extent, for numerical models in computational astrophysics. Online science platforms with astronomical data have been made available in the past few decades, e.g. the \href{https://www.sdss.org/data-releases}{Sloan Digital Sky Survey} \citep{2000AJ....120.1579Y}, the \href{https://www.darkenergysurvey.org/}{Dark Energy Survey Science Portal} \citep{2005astro.ph.10346T, 2018A&C....24...52F} to provide the scientific community with access to theses surveys. In the same manner, similar dedicated science portals are currently being developed for observational data that will be produced by the Euclid space telescope \citep{2011arXiv1110.3193L}, the Square Kilometer Array (SKA) telescope \citep{SKA2009} or the Vera Rubin Observatory \citep{LSST2019}.

If the dissemination of astronomical data has been greatly facilitated by the standardization of file format, for example the FITS format \citep{FITS1981}, or the MeasurementSet \citep{MeasurementSet2.0_2000} standard in radio astronomy, computational astrophysicists lack a standardized data model to enable them to exchange their numerical simulation data in a uniform way. Nevertheless, a few simulation projects have released their data to the community, many of them in the cosmology field: e.g. the MultiDark \citep{Prada2012_MultiDark} and Bolshoi \citep{Klypin2011_Bolshoi} simulations hosted on the \href{https://www.cosmosim.org}{CosmoSim} database, the \href{http://gcmc.hub.yt}{galaxy cluster merger catalog} \citep{2018ApJS..234....4Z} hosted on the \href{http://hub.yt}{yt Hub}, the \href{https://www.illustris-project.org}{Illustris project} \citep{2014MNRAS.444.1518V, 2019ComAC...6....2N},
the  \href{https://c2papcosmosim.srv.lrz.de}{web portal for cosmological hydrodynamical simulations} \citep{2017A&C....20...52R},
\href{https://www.cosmohub.pic.es}{CosmoHub} \citep{2017ehep.confE.488C, 2020A&C....3200391T} or the \href{https://tao.asvo.org.au/tao}{Theoretical Astrophysical Observatory} \citep{2016ApJS..223....9B}. Some projects even attempted to combine infrastructures in the field of turbulence studies, e.g. the \href{https://turbulence.pha.jhu.edu/}{Johns Hopkins Turbulence Database} \citep{2008JTurb...9...31L}, the \href{https://www.mhdturbulence.com}{Catalogue for Astrophysical Turbulence Simulations} \citep{2020ApJ...905...14B} or in the field of star formation (the \href{http://starformat.obspm.fr}{StarFormat database}) or interstellar medium (\href{http://ismdb.obspm.fr}{ISMDB}).

One of the major collaborative efforts led to the release of the generic \href{https://django-daiquiri.github.io}{Django-Daiquiri framework} \citep[and references therin]{2020ASPC..527..395G}, a web application developed with the aim of being deployed for each project to host the data produced by the numerical simulations conducted in the project.
It greatly lowered the technical barrier individual research groups have to overcome to publish their data on a web application, but unfortunately the required technical expertise and maintenance expenses are still out of reach for a majority of small individual projects led by computational astrophysicists.

\section{The Galactica simulation database}\label{sec_galactica}

\begin{figure}[h]
    \centering
    \includegraphics[width=\textwidth]{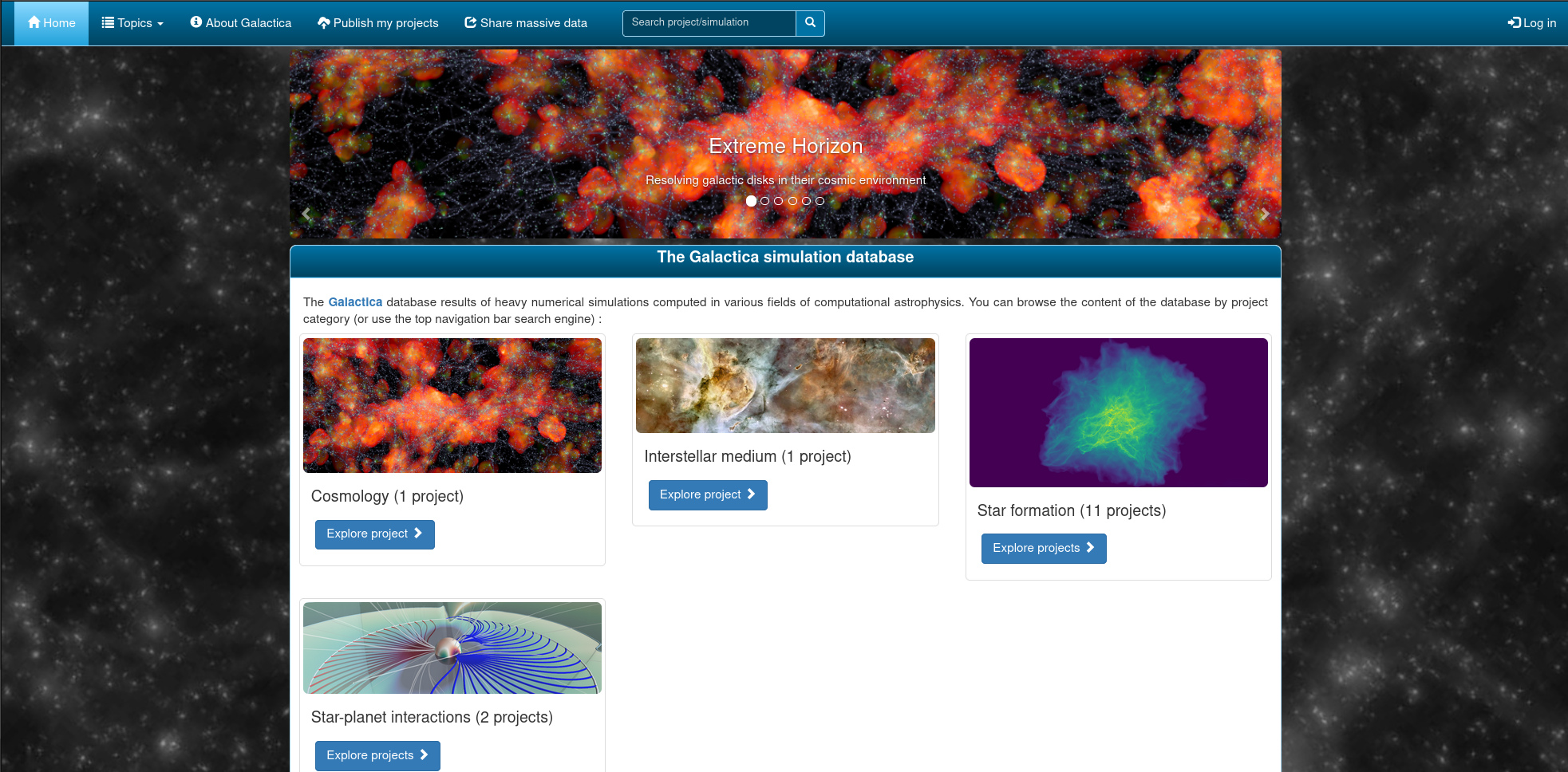}
    \caption{\href{http://www.galactica-simulations.eu}{Galactica database home page}, with public projects sorted by categories.}
    \label{fig:galactica_home}
\end{figure}

The \href{http://www.galactica-simulations.eu}{Galactica simulation database} has been developed with both generality and versatility in mind. The web application datamodel is an implementation of the \href{https://www.ivoa.net/documents/SimDM}{SimulationDatamodel IVOA standard}, using a classical MySQL relational database as a backend to store metadata related to hosted scientific projects. This \href{https://ivoa.net/}{IVOA} standard is completely generic and enables the description of any type of numerical simulations. As a consequence, Galactica can technically host any scientific project (see Fig. \ref{fig:galactica_home}) in computational astrophysics, from any field of research (cosmology, galaxy formation, interstellar medium, solar physics, planetary atmospheres, star-planet interactions, star formation, etc.). Hence, the web application is a collaborative platform where all scientists of the community can request an account, publish their project or access available simulation data.

Contrary to the Daiquiri approach, the Galactica database is a unique deployment centralising all the metadata of different computational astrophysics projects, where scientists can edit their own project pages through a \textit{content management system} (CMS). They have the possibility to document their project, their simulations, the code they used, the implementation details, the numerical setup and  they can publish their reduced datasets (e.g., plots, images, spectra, object catalogs) directly online. The catalogs can be accessed via search forms and reduced datasets can be visualized via visualization components \citep[e.g., \href{https://leafletjs.com/}{LeafletJS}, \href{https://d3js.org/}{D3JS}:][]{Bostock_D3JS2011}. The lack of project specific online functionalities have been mitigated by the integration of many general purpose functionalities that could benefit all kinds of projects hosted on the platform.

In section \ref{sec_terminus}, we describe how to disseminate and access raw data generated by massive simulations in astrophysics through Galactica. In section \ref{sec_astrophysix}, we present a Python API for Galactica to ease and automate the process of data ingestion on the web application for computational astrophysicists.

\section{Distributed ecosystem of raw data-processing servers}\label{sec_terminus}

\begin{wrapfigure}{l}{0.6\textwidth}
    \centering
    \includegraphics[width=0.6\textwidth]{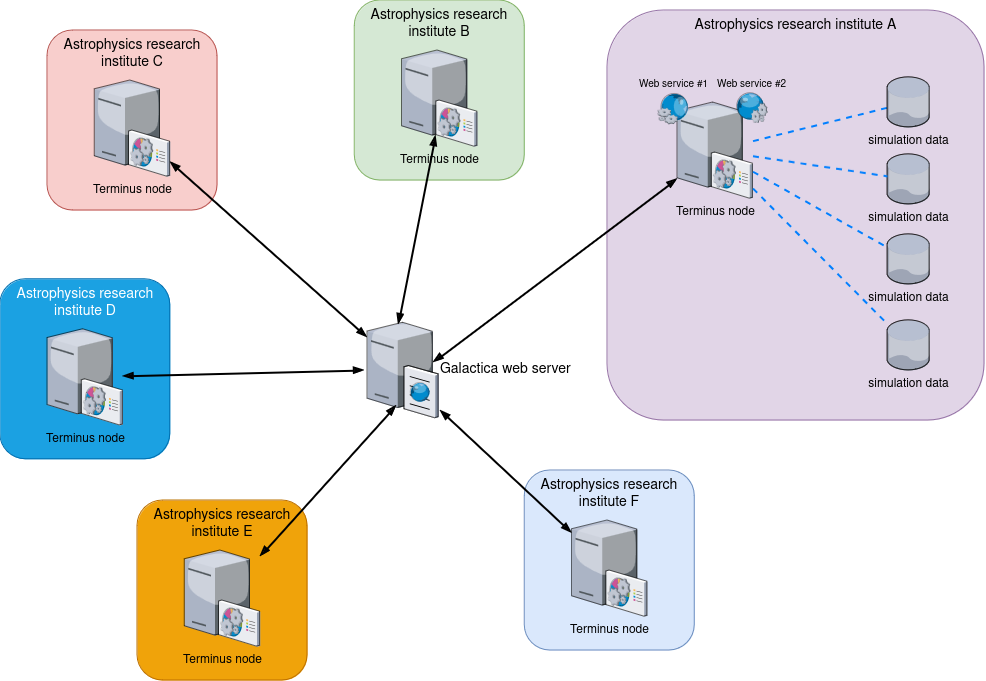}
    \caption{Galactica database distributed data-processing architecture, with multiple \textit{Terminus} nodes in different research institutes or supercomputing facilities.}
    \label{fig_distrib}
\end{wrapfigure}

Reduced datasets are both easier and cheaper to transfer, store, index and publish due to their reasonably low volumes, which makes them perfectly adapted for dedicated open science platforms like \href{https://zenodo.org}{Zenodo}, or the \href{https://research-and-innovation.ec.europa.eu/strategy/strategy-2020-2024/our-digital-future/open-science/european-open-science-cloud-eosc_en}{European Open Science Cloud}. But a technical solution is still missing to provide access to the massive simulation datasets produced by numerical codes, with volumes on the order of gigabyte (GB) or terabyte (TB) or even petabyte( PB). Galactica provides a solution by interfacing with a distributed ecosystem of raw simulation data processing servers, called \textit{Terminus} nodes. We developed the \href{https://pypi.org/project/galactica-terminus}{\textit{Terminus}}\footnote{\textit{Terminus} online documentation: \url{https://galactica-terminus.readthedocs.io}} server as a custom overlay on the \href{https://docs.celeryq.de}{Celery} asynchronous task management framework. Each \textit{Terminus} node provides storage for massive simulation datasets and computational resources to process them and generate high-level data products that fit the custom needs of the scientific community. Computational astrophysicists willing to give access to their raw simulation data can store them on a local \textit{Terminus} node hosted by their research institute or a supercomputing facility (see Fig. \ref{fig_distrib}). In addition, they are also responsible for deploying and maintaining the data-processing services on their local \textit{Terminus} server, with the software tools of their choice compatible with the data format produced by the simulation code.

This distributed architecture is not only cost efficient but also very scalable and guarantees excellent sovereignty over the data, the storage and computational resources contributed to the Galactica ecosystem. While the Galactica centralised web application is storing lightweight project-related metadata and reasonably lightweight reduced datasets, it only requires a small storage capacity, which makes it inexpensive. The highly scalable distributed network of \textit{Terminus} nodes provides virtually all the computing power and storage capacity, in line with local research institute policy. Galactica opens the possibility for \textit{Terminus} node hosting institutes to apply for specific funding for these open science dedicated resources.

Individual projects can keep complete sovereignty over the data they want to share with the community (in agreement with their project \textit{Data Management Plan}), and the high-level data products that can be generated from it. The \textit{Terminus} implementation makes it compatible with any type of data processing service, scientific analysis library and simulation data format, which contributes to the very high versatility of the infrastructure. To achieve the scientific community requirements, these data products would ideally be 1D profiles, 2D projections, 3D datacubes, images, spectra or any kind of subsampled datasets, generated in a standard file format (e.g. FITS, HDF5, PNG, CSV, JSON). These services are accessible online through Galactica via "on-demand" job request forms with as many parameterization options as needed.

\section{Automated offline project documentation with \textit{astrophysix}}\label{sec_astrophysix}

The Galactica online content management system could prove tedious to use for physicists who would need to upload a lot of metadata on their project pages by hand. This is why the \href{https://pypi.org/project/astrophysix}{astrophysix}\footnote{\textit{astrophysix} online documentation : \url{https://astrophysix.readthedocs.io}} Python package  has been developed alongside the Galactica web application to let computational astrophysicists automate their numerical project documentation with an API that can be integrated in an existing Python scientific analysis pipeline. This API also implements the \textit{SimulationDatamodel} IVOA standard and is intended to be used to document all the meta-information related to the simulation project and attach the reduced datasets obtained during the scientific analysis.

\begin{lstlisting}[language=Python, caption=Project documentation with the astrophysix Python API, label=lst:astrophysix_example]
from astrophysix.simdm import SimulationStudy, Project,
                              ProjectCategory

proj = Project(category=ProjectCategory.Cosmology,
               alias="EXT_HORIZON")
study = SimulationStudy(project=proj)
study.save_HDF5("./EH_study.h5") # to upload on Galactica
\end{lstlisting}


 Once documented, a project can be exported as a standard HDF5 file and uploaded on the Galactica web application to deploy or update all the project pages instantaneously (see Listing \ref{lst:astrophysix_example}). For each project, this can be done by several contributors to fill the project page content within a scientific collaboration.

\section{Conclusion}\label{sec_conclusion}

The data publication and reuse opportunities brought by the Galactica infrastructure can be used effectively for theory validation, numerical model comparison, observational campaign preparation or comparison with already available astronomical data. 

The platform offers free data access and is open to contributions. Together with the \textit{Terminus} server and the astrophysix API, they form a Galactica software suite that can be useful for the vast majority of computational projects, with small resources, and no advanced technical expertise required to publish and disseminate simulation data in astrophysics. Although it delegates massive simulation data storage and processing to remote \textit{Terminus} servers, where data-processing services must be maintained by the hosting institutes, its good generality, scalability and versatility makes it a powerful tool for open science in computational astrophysics.

\bibliographystyle{aa}
\bibliography{galactica.bib}




\end{document}